\documentclass{emulateapj}

\usepackage{float}
\usepackage{amsmath}
\usepackage{epsfig,floatflt}
\usepackage{subfigure}

\begin{document}

\title{The Cosmological Mean Density and its Local Variations Probed by 
Peculiar Velocities}

\author{Roya Mohayaee\altaffilmark{1}, R. Brent Tully\altaffilmark{2}} 
\altaffiltext{1}{Institut d'astrophysique de Paris, 98bis bd Arago, 75014 Paris, France}

\altaffiltext{2}{Institute for Astronomy, University of Hawaii, Honolulu, HI 96822, USA}

\begin{abstract}
Peculiar velocities throughout the region of the Local Supercluster are 
reconstructed by two different orbit-retracing methods. 
The requirement of the optimal correlation between the radial components of
reconstructed velocities and the observed peculiar velocities derived from
our extensive new catalog of distances puts 
stringent constraints on the values of the cosmological parameters.
Our constraints intersect those from studies of microwave background
fluctuations and statistical properties of galaxy
clustering: the ensemble of constraints are 
consistent with $\Omega_m=0.22\pm0.02$.
While motions throughout the Local Supercluster provide a
measure of the mean ratio of mass to light, there 
can be large local fluctuations.
Our reconstruction of the infall velocities in the immediate
vicinity of the Virgo Cluster shows that there is a
mass-to-light anomaly of a factor of 3 to 6 between groups in the
general field environment and the heavily populated Virgo Cluster.

\end{abstract}


\maketitle

\section{Introduction}
\label{sec:introduction}

Cosmological parameters have
been significantly constrained by the combined analyses of 
the data mainly from Cosmic Microwave Background (CMB) radiation,
large-scale galaxy distribution, supernova observations and
cosmic shear field [see {\it e.g.} \citep{teg04}].
Here, we demonstrate that analysis of the peculiar velocity field 
can tighten the constraints on cosmological models because error ellipses 
are strongly skewed
to those of {\small \emph{WMAP}} and {\small SDSS} in 
the parameter space of density and age.
Two complementary theoretical methods 
are used in this paper to generate
orbits that give velocities to be compared with observations.  
Peculiar velocities are defined
by a large number of quality distance measures.  Good distances are 
available for 1400 galaxies within 3,000~km~s$^{-1}$.

The two variational methods used here are based on the principle that 
galaxies as mass-tracers follow
orbits which are the stationary points of the Euler-Lagrange action. 
Our first approach, the {\it Least Action} ({\small LA}) method \citep{pee89},
describes orbits in the nonlinear collapse regime,
though recovered trajectories are non-unique due to lack of knowledge 
of the initial positions of 
galaxies and because of orbit-crossing ({\it i.e.} multisteaming). 
On large scales, a unique
reconstruction is possible since the displacement of dark matter fluid 
derives from a potential and multistreaming is less severe. 
These properties are at the basis of our second method, {\small MAK} 
\citep{fri02}. The methods are applied at the Local Supercluster scale for the
determination of the cosmological mean density and at the cluster scale 
for the specification of its local variations.

Allowing for statistical and systematic errors and considering all
the methods
in tandem, we show that the {\small MAK}, {\small LA}, {\small \emph{WMAP}}
and {\small SDSS} results
give preference to a
small region in parameter space favouring a low value of 
$\Omega_m=0.22\pm 0.02$ for the density parameter. 

On cluster scales, 
reconstruction of the amplitude of the first-approach 
infall velocities provides a determination of the mass of the
clusters at radii beyond the virialized region; radii greater than those
explored by X-ray or lensing studies.
We show that the high amplitude observed in the infall region
requires an assignment of
$M/L=500 M_{\odot}/L_{\odot}$ to the Virgo Cluster 
(where the luminosities are in the B band).

\section{Catalog of Nearby Galaxies and distances}
\label{sec:catalogs}

The catalog of galaxies, is a
40\% augmentation of the Nearby
Galaxies Catalog \citep{tul87}, now including 3300 galaxies within
3,000~km~s$^{-1}$.
This depth is more
than twice the distance of the dominant component, the Virgo Cluster, 
and the completion
to this depth in the current catalog compares favorably with
other all-sky surveys
[{\it e.g.}, 2MASS \citep{jar04}].
The catalog has the following properties.
(i) All entries are given a detailed
group and filament assignment (or non-assignment if isolated).
(ii) The catalog is supplemented by an all-sky complete sample of 
X-ray selected clusters \citep{koc04} in the shell 3,000--8,000 km~s$^{-1}$ to provide
a description of potential influences on very large scales.
(iii) `Fake' galaxies are added at low Galactic latitudes to avoid an
underdensity in the zone of obscuration due to lost information.
(iv) Correction is made for the loss of light with distance caused by
an apparent magnitude cutoff in the construction of the catalog.

The second observational component is an extended catalog of galaxy 
distances.  Information from four techniques has been
integrated: the Cepheid variable \citep{fre01}, Tip of the Red Giant 
Branch \citep{kar03,lee93}, Surface Brightness Fluctuation \citep{ton88,ton01}, and 
Luminosity--Linewidth \citep{tul77,tul00} methods.
In all, there are over 1400 galaxies with
distance measures within the 3,000~km~s$^{-1}$ volume; over 400 of
these are derived by at least one of the first three `high quality'
techniques.  In the present study, distances are averaged over groups
because orbits cannot meaningfully be recovered on sub-group scales.
The present catalog is assembled into 1234 groups (including groups
of one) of which 633 have measured distances.
These observational components will be described in detail in a later
publication.

\section{Techniques: the LA and MAK methods}
\label{sec:techniques}

Peebles pioneered the Least Action (LA) method \citep{pee89,pee94,pee95} that requires
orbits to satisfy the stationary point of the Euler-Lagrange action, the
integral over time of 
the Lagrangian. Hence, least action searches for the minimum of
\begin{equation}
\!\!S_{\rm LA}\!=
\!\!\!\int_{0}^{t_0}\!\!\!\!\! dt\left[
{m_i a^2 \dot{\bf x}_i^2\over 2}
-{G m_i m_j\over a\vert {\bf x}_i-{\bf x}_j\vert}
+{2\over 3}\pi G\rho_{\rm b} a^2 m_i {\bf x}_i^2\right],\!\!
\label{el-action}
\end{equation}
where summation over repeated indices and $j\not=i$ 
is implied, $t_0$ denotes the present time, 
the path of the $i$th particle with mass $m_i$ is ${\bf x}_i(t)$, 
$\rho_{\rm b}$ is the mean mass density, 
and the present value of the expansion
parameter $a(t)$ is $a_0=a(t_0)=1$.

Individual orbits are constrained 
by the mixed boundary conditions
that peculiar velocities were initially negligible (they subsequently
grew from gravitational perturbations) and the known
elements of position and velocity today. 
The other piece of information known for roughly half the
elements is the aforementioned distances. This information is used to discriminate
between models. A given model involves a specification of
cosmological parameters and the assignment of mass to each of the
elements.  Orbits are found within the context of a specified model
that are consistent with the boundary condition constraints on angular
positions and radial velocities. A component of the end point of an
orbit is its distance which can be compared with the observed
distance. Different models result in different distances.  The
quality of a model can be evaluated by a statistical measure of the
differences between observed and model distance moduli,
$\mu_{obs}$ and $\mu_{mod}$ respectively. A $\chi^2$ parameter can be
calculated for each element, $i$, with a measured distance,
$\chi^2_i = (\mu_{mod,i} - \mu_{obs,i})^2 /  \epsilon_i^2$,
where $\epsilon_i$ is the uncertainty assigned to $\mu_{obs,i}$.
A comparison of the relative distribution of $\chi^2_i$ values
provides a sensitive discriminant between good and bad models \citep{sha95,phe02}.

Galaxy flows are not sensitive to any 
uniform repulsive dark energy \citep{lah91,sha95}.  
Measurements of the lowest frequency peak in the microwave background
fluctuation spectrum strongly indicate, and we accept, that the Universe 
has a flat topology,
whence $\Omega_{\Lambda} + \Omega_m = 1$ where  $\Omega_{\Lambda}$
characterizes the mean density of dark energy and $\Omega_m$ characterizes
the mean density of matter with respect to the critical density for a 
flat Universe in matter alone.
Specification of $\Omega_m$ and the Hubble Constant, H$_0$, 
which describes the mean expansion rate of the Universe, in
the context of a flat Universe uniquely specifies the age of the
Universe, $t_0(\Omega_m,{\rm H}_0)$.

The Least Action orbits are a reflection of the interactions of masses
over the age of the Universe.  As an initial approximation, it is assumed that
the relationship between mass and light is constant so the mass $M_i$
assigned to an element in the catalog with luminosity $L_i$ is
$M_i=(M/L)L_i$.  Consequently, a cosmological model to be explored is
specified by two fundamental parameters: $t_0(\Omega_m,{\rm H}_0)$ and
$M/L$.  For a given choice of $t_0$ and $M/L$, orbits are constructed, 
model distances at
the end points are determined and compared with observed distances,
and the match provides a $\chi^2$ measure that can be compared
with alternative models.  Discrimination between results over the
domain of ($t_0, M/L)$ choices based on the $\chi^2$ measure gives
specification of a best model.

The alternate technique that will be implemented here
is the Monge-Amp\`ere-Kantorovich ({\small MAK}) reconstruction \citep{fri02,moh03,bre03}.
This method provides a recipe for the orbits
that is unique to the degree that orbits can be described as following
straight lines under suitable coordinate transformations. 
Orbits in the {\small MAK} reconstruction are
minima of the action which assume a Lagrangian mapping ${\bf q \mapsto x}$
that can be described in terms of a potential as
${\bf x=\bigtriangledown_q \Phi(q)}$.
Here, ${\bf x}$ is the current Eulerian position and ${\bf q}$
is the initial Lagrangian position.
The potential ${\bf \Phi(q)}$
is assumed to be convex; ie, orbit-crossing is excluded. With these
conditions, orbits are reconstructed uniquely by an assignment
algorithm that finds the minimum in
the ensemble of orbit distances summed in quadrature.
 Hence, {\small MAK} search for the minimum of
the discretized action
\begin{equation}
S_{\rm MAK}=\sum_{i=1}^N\left({\bf q}_{j(i)}-{\bf x}_i\right)^2 .
\label{mak-action}
\end{equation}
Initial positions ${\bf q}_j$, are assigned to final 
positions ${\bf x}_i$. For each final position $i$, there are $N$ 
possible initial 
positions and only the one that minimizes the sum in the expression (2) 
is allowed.   A brute-force
assignment algorithm would have a complexity of $N!$, but can be
reduce to a polynomial complexity (specifically, so far to $N^{2.25}$).
It is insightful to note that the least action variation (\ref{el-action})
reduces to {\small MAK} optimization (\ref{mak-action})
for inertial trajectories 
({\it i.e.} where particles move with their initial velocities).

The distances, $d$, permit an extraction of
peculiar velocities $V_{pec} = V_{gsr} - d {\rm H}_{0}$ where $V_{gsr}$ is
the observed velocity of an object in the galactic standard of rest.
The relationship between redshift and real space 
displacements of the elements is
estimated using the approximation that galaxies trace
mass on inertial trajectories \citep{zel70}:  
${\bf v} = f{\rm H}({\bf x} - {\bf q})$ where ${\bf v}$ is the peculiar
velocity vector and $f \sim \Omega_m^{0.6}$.
Consequently, a specific model defines positions that can be tested against
observed positions in the same way that is done in the Least Action
analysis.

An advantage of Least Action is that it potentially
can be used in the highly non-linear regime where orbits are strongly
curved. An advantage of {\small MAK} is that it recovers orbits uniquely and quickly, hence
can be used with the very large catalogs that are becoming
available.  
By using the two methods on the identical catalogs, with
the identical distance constraints, it is affirmed that the two
methods recover similar results in the regimes where they are both
applicable.

Both methods have been tested on N-body simulations and found to result
in modest underestimations of the known densities. Underestimates are
anticipated because, at one extreme, mass that is not
strongly clustered on the scale of the survey has little dynamical 
consequence and, at the other extreme, complex shell-crossing orbits 
cannot be modeled.
Analysis of {\small MAK} reconstructions in 12 cosmological simulations with different
initial fluctuation characteristics discussed by
\cite{moh05} resulted in recovery of $80\% \pm 20\%$ ($1 \sigma$) of 
the known model density.  In the case of Least Action, it has been 
appreciated that mass can be underestimated on small scales 
\citep{bra94,bra02}. A quantitative measure of the effect is found using
our Least Action reconstruction algorithms with
an N-body simulation with $\Omega_\Lambda=0.7$, $\Omega_m=0.3$,
evaluated by
placing the observer at multiple locations to test for cosmic variance.
Recovered densities were $70\% \pm 20\%$ of the model
density (Phelps, Desjacques, and Nusser; ongoing work).
The estimates that we quote for $\Omega_m$ include adjustments for the 
systematics of 20\% and 30\% for MAK and Least Action,
respectively.

\section{Results: I. determination of $\Omega_m$}
\label{sec:result1}

\begin{figure}
\begin{center}
\mbox{\epsfig{figure=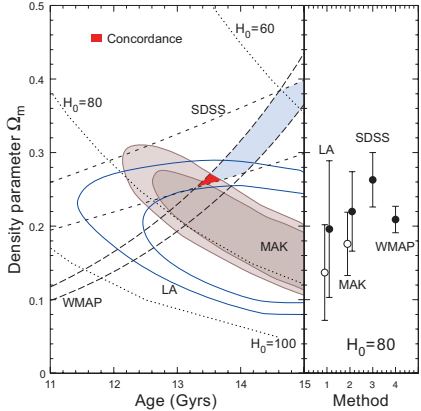,width=0.646\linewidth,clip=}}
\end{center}
\vspace*{-0.4cm}
\caption{
\noindent
Reconstruction methods constrained by precision distance measures 
give estimates of the mean mass density of the Universe.
The contours are $\chi^2$ at 1 and $2 \sigma$ for models with different 
choices of age and mass density $\Omega_m$, with shaded contours for {\small MAK} and 
line contours for {\small LA} reconstructions. 
The {\small MAK} and Least Action contours are adjusted upward in density by 20\% and
30\% respectively to compensate for
systematic underestimations of density.
The confluence of the constraints on density and age parameters
from {\small \emph{WMAP}} \citep{spe03} of $\Omega_m h^2=0.134$ 
and from {\small SDSS} \citep{teg04} of $\Omega_m h=0.21$ is
lightly shaded.
The darkest shaded region
lies within the $2 \sigma$ limits of all four experiments.
The right panel shows the 
density determinations for {\small LA}, {\small MAK}, {\small SDSS}, and
{\small \emph{WMAP}} 
for H$_0 = 80$~km~s$^{-1}$~Mpc$^{-1}$, the value of the Hubble Constant 
consistent with 
our distance scale zero point. The open and filled symbols for {\small LA} and
{\small MAK}
respectively give the results before and after compensation for 
the systematic under-determination of mass density. At H$_0 = 80$, 
$t_0^{SDSS}=12.2$~Gyr,
$t_0^{WMAP}=13.0$~Gyr, and $t_0^{MAK,LA}=13.5$~Gyr.
The 4 analyses are consistent with 
$\Omega_m = 0.22 \pm 0.02$ with the constrained value of H$_0$.
\label{fig1}}
\end{figure}

The results of the two methods are summarized in
Fig.~1 where contours of the $\chi^2$ parameter are plotted in the
domain \{$t_0$,$\Omega_{\rm m}$\}.
In the case of the Least Action analysis, mass density $\Omega_m$ is
derived from $M/L$ values 
by accepting the mean $B$ band luminosity density of the Universe 
determined from {\small SDSS} \cite{bla03} with reddening corrections.  
We use $\Omega_m = (M/L) / (1540 h)$, a transform with $\sim 20\%$ uncertainty.

The error ellipses from the {\small MAK} and {\small LA} studies are elongated.  
With a linear analysis the elliptical $\chi^2$ troughs would open to
infinity, but non-linear effects create a specific minimum along each 
$\chi^2$ trough.
At a fixed age, $t_0$, there are relatively tight constraints
on $\Omega_m$ requirements. If shorter ages are entertained,
then higher densities are required to arrive at the observed dynamical state
in the specified time.  The contrary dependence between density and time
are seen in the 
results from the microwave background {\small WMAP} \citep{spe03} and galaxy redshift 
{\small SDSS} \citep{teg04} experiments that are also superimposed
on Fig.~1.  
Only a small domain around $t_0=13.5$~Gyr 
lies within or near the $2 \sigma$ contours of all the methodologies.
With the constraint $h=0.8$ consistent with the zero point of the
distance estimates, there is good agreement between the {\small WMAP}, {\small
SDSS}, {\small MAK},
and {\small LA} measures of the density parameter: $\Omega_m=0.22\pm0.02$.
The low value of $\Omega_{\rm m}$ obtained here is consistent with 
that found from divers recent studies 
\citep{ostriker20031, ostriker20032, vdb03, col05}.

\section{Results: II. the mass of Virgo Cluster}
\label{sec:result2}

The velocity field analyses discussed above make the simplistic assumption
of constant $M/L$.  
There has been evidence in the literature, though, for large variations in 
$M/L$ with environment \citep{mar02, vdb03, eke04,
tul05b, vdb05, par05}.
Here we note that galaxy flows near the Virgo
Cluster contradict the hypothesis of constant $M/L$.
The high velocities of infall suggest a much larger $M/L$
for the cluster than the value consistent with a good fit over the
full region within 3,000~km~s$^{-1}$.
The critical galaxies for this discussion lie outside the cluster
on the plane of the sky so only a modest component of the
infall motion is projected into the line of sight but there is no
confusion with cluster membership.  Galaxies infalling
from the foreground of the cluster are redshifted with respect to the
cluster and galaxies infalling from the background are blueshifted.
If one could imagine many test particles distributed along a
line of sight, they would create a wave in velocity as a function of
distance with
the same velocity arising at three distances \citep{ton81}
-- two located within the infall region and one located at the
cosmic expansion position.
The amplitude of the peaks of the `triple-value' waves depend on the
mass, $M$, interior to the position associated with the peaks, $r$.
The envelope of observed infall velocities with $r$ provides a 
description \citep{tul84} of the run of $M(<r)$.
The outermost caustic of the Virgo Cluster
proper (the radius of second turnaround) is at $\sim 2$~Mpc from the
center of the
cluster.  Galaxies on first infall acquire very high radial
velocities at 2--4 Mpc from the Virgo core.

\begin{figure}
\begin{center}
\mbox{\epsfig{figure=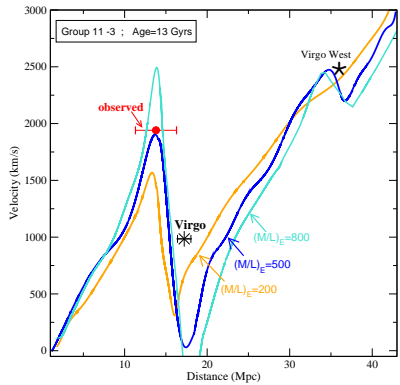,width=0.615\linewidth,clip=}}
\end{center}
\vspace*{-0.4cm}
\caption{
The distortion of the velocity--distance relation by the Virgo
Cluster in the line of sight toward the 11--3 group.  
The observed distance
and velocity of the 11--3 group is indicated by the point with error bars
in distance (errors in velocity are smaller than the symbol).
The observed distance and velocity of the Virgo Cluster is indicated
(centered at 6.2 degrees from the line of sight of the 11--3 group).
The background Virgo W Cluster causes the second wiggle along this line
of sight.
The behavior of
infall into the Virgo Cluster is traced by the locus of positions and
velocities of test particles of negligible mass that were laid down
in the line of sight of the 11--3 group and subjected to Least Action
modeling. 
It is concluded from the information seen on this plot and equivalent plots of the
lines of sight to other groups close to the Virgo Cluster that $M/L_{\rm
Virgo}\approx 500 M_{\odot}/L_{\odot}$ and the mean $M/L$ for
the field outside the cluster is approximately $1/3$ of this value. 
\label{fig2}
}
\end{figure}

An example is given in Fig.~2.  The run of
velocities with distance are what is expected along the specific
line of sight of the 11--3 group in the Nearby Galaxies Catalog
with three different assumptions regarding the mass of the Virgo Cluster.
The observed distance and velocity of the 11--3 group is indicated in
the figure. It is clear that the 11--3 group is at one of the two
infall locations (or the location where the two infall positions become
degenerate at the tip of the triple-value curve) rather than at the cosmic
expansion position around 30 Mpc.  

The total mass within 3,000~km~s$^{-1}$ is the same in all 3 models
to ensure that the global $\chi^2$ is near the minimum of the trough in
Fig.~1, however the distribution of mass between the elliptical rich and spiral
rich groups is modified between models.  Here, $t_0=13$~Gyr.
In the case of the curve with
the smallest swing, all components are given the same
$M/L=200 M_{\odot}/L_{\odot}$.  The intermediate curve is generated with
the elliptical-dominated groups (including the Virgo Cluster) given
$M/L_E=500$ and the spiral-dominated groups given $M/L_S=174$.  The curve with the
largest swing reflects $M/L_E=800$ and $M/L_S=148$.  
In the model illustrated by the middle curve in
Fig.~2, the cluster is given the necessary and just sufficient mass of
$9 \times 10^{14} M_{\odot}$.  

The 11--3 group and a couple of others provide the greatest demands
on the mass of the cluster and are inferred to lie near a tip along a
triple-value curve.
An adequate description of the amplitude of their
infall velocities requires an assignment of
$M/L=500 M_{\odot}/L_{\odot}$ to the Virgo Cluster if
$t_0=13$~Gyr.  As with the overall Supercluster modeling, less cluster
mass is required if given more time \citep{tul84}: {\it e.g.}, 
$8 \times 10^{14} M_{\odot}$
suffices if $t_0=14$~Gyr.  In both these 13 and 14 Gyr cases,
as the mass assigned to the cluster is augmented,
the models
require a reduction of mass assigned to the field to remain at the
minimum of the $\chi^2$ trough of Fig.~1.
The ratio of cluster, $C$, to field, $F$, mass to light values is
$(M/L)_C/(M/L)_F \simeq 3$
with the mass assignments required to explain the
infall motions.  This factor 3 probably underestimates the $M/L$ difference
between Virgo and bound groups in the field because some of the mass
contributing to the field ratio lies outside the groups.  
Studies of the dynamics of nearby groups \citep{tul05b} suggest that bound
groups of spiral galaxies in the field have $M/L \sim 90$.  The mass to
light ratio of the Virgo Cluster is 5--6 times larger.

%\vspace*{-0.5cm}
\section{Conclusions}
\label{sec:conclusion}

Our first result on the mean density of the Universe 
complements the {\small \emph{WMAP}} and {\small SDSS} measurements of
the density parameter because the error 
ellipses in the domain $\Omega_m, H_0$ are 
steeply inclined to each other. A narrow 
range of parameter space around $t_0=13.5$ Gyr,
$\Omega_m=0.22\pm0.02$ is permitted by the combined experiments.  

Our second result concerns the local fluctuations around this mean density.
Although most of the blue light in the local Universe is
associated with the spiral field, there is close to parity in the
partition of mass between the spiral field and the Virgo Cluster.
The $M/L$ value associated with the E/S0 dominated Virgo
Cluster is significantly higher than that associated with the spiral
dominated field.  The complex problem of $M/L$ variations with environment
shall be discussed in a future more-detailed paper.  

\smallskip

% Acknowledegement

{\small
We thank Ed. Bertschinger, J. Colin, U. Frisch, M. H\'enon, G. Lavaux, 
P.J.E. Peebles, S. Phelps, E. Shaya and J. Silk for 
discussions, collaboration and support.
RM was supported by a Marie Curie fellowship. RBT acknowledges
support from the BQR program of the Observatoire de la C\^ote d'Azur,
the French Hautes Niveau of EGIDE, a JPL Space Interferometry Mission contract, 
and STScI and NSF awards.
}

\end{document}